\documentclass[twocolumn,prb,aps,superbib,tightenlines,floatfix]{revtex4}
\usepackage{amsfonts,amsmath,amssymb,amsthm,bm}
\usepackage{graphicx}
\begin{document}
\bibliographystyle{apsrev}
\title{Atomic-scale Field-effect Transistor as a Thermoelectric Power Generator and Self-powered Device.}
\author{Yu-Shen Liu }
\author{Hsuan-Te Yao}
\author{Yu-Chang Chen}
\email{yuchangchen@mail.nctu.edu.tw}
\affiliation{Department of Electrophysics, National Chiao Tung University, 1001 Ta Hush
Road, Hsinchu 30010, Taiwan}
\begin{abstract}

Using first-principles approaches, we have investigated the thermoelectric
properties and the energy conversion efficiency of the paired metal-Br-Al
junction. Owing to the narrow states in the vicinity of
the chemical potential, the nanojunction has large Seebeck coefficients such
that it can be considered an efficient thermoelectric power generator.
We also consider the nanojunction in a three-terminal geometry, where
the current, voltage, power, and efficiency can be efficiently modulated by
the gate voltages. Such current-voltage characteristics could be useful in the design
of nano-scale electronic devices such, as a transistor or switch.
Notably, the nanojunction as a transistor with a fixed finite
temperature difference between electrodes can power itself using the Seebeck effect.
\end{abstract}
\pacs{73.50.Lw, 68.43.Pq, 73.40.Jn 81.07.Nb}
\maketitle

\section{Introduction}

In the past decade, considerable concern has arisen regarding the transport
properties of atomic-scale junctions, which are the basic building blocks for
molecular electronics~\cite{Aviram,Reed1}. This concern is motivated by the
aspiration to develop new forms of electronic devices based on subminiature
structures and by the desire to understand the fundamental properties of
electron transport under non-equilibrium~\cite{book}. A growing number of
research studies are now available to diversify the scope of electron
transport properties in molecular/atomic junctions, including current-voltage
characteristics~\cite{Kaun,DiVentra2,Nitzan1}, inelastic electron tunneling
spectroscopy
(IETS)~\cite{Wang2,Ratner3,Yluo,Kushmerick,Brandbyge,Dicarlo,alkane,cheniets,chenh2,thygesen}%
, shot noise~\cite{Ruitenbeek1,Kiguchi,Natelson,chenshot}, counting
statistics~\cite{liushot}, local heating~\cite{chenheating,Huang2}, and
gate-controlled effects~\cite{DiVentragate,Ma,Solomon1,Solomon2,Reed}.
Substantial progress in experiment and theory has been
achieved~\cite{Ahn,Lindsay,Tao1}.

Recently, growing attention has been paid to the thermopower of nanojunctions.
Pioneering experiments have been conducted to measure the Seebeck coefficients
at the atomic and molecular level~\cite{Ruitenbeek,Reddy1,Reddy2,Reddy3}. The
Seebeck coefficient is related not only to the magnitude, but also to the
slope of the transmission function in the vicinity of chemical potentials.
Thus, it can provide richer information than the current-voltage
characteristics regarding the electronic structures of the molecule bridging
the electrodes~\cite{Malen}. The thermoelectric effect hybridizes the electron
and energy transport, which complicates the fundamental understanding for
quantum transport of electron and energy under non-equilibrium conditions.
This has spurred rapid developments in the fundamental thermoelectric theory
in nanojunctions~\cite{Paulsson,Zheng,Wang,Pauly,Dubi,Markussen,Ke,Finch,Troels,Bergfield,Liu2}
including the effects of electron-vibration interactions
~\cite{Galperin,Imry,Hsu}. The Seebeck coefficient is typically measured
under equilibrium condition, non-equilibrium current and inelastic effects
potentially offer new possibilities of engineering systems leading to enhanced
the thermopower~\cite{Esposito,Liu}. The emergence of thermoelectric
nanojunctions may also have profound implications on the design of
subminiature energy-conversion devices, such as
nano-refrigerators~\cite{DiventraS,Galperin2,Liu3}.

\begin{figure}[ptb]
\includegraphics[width=8cm]{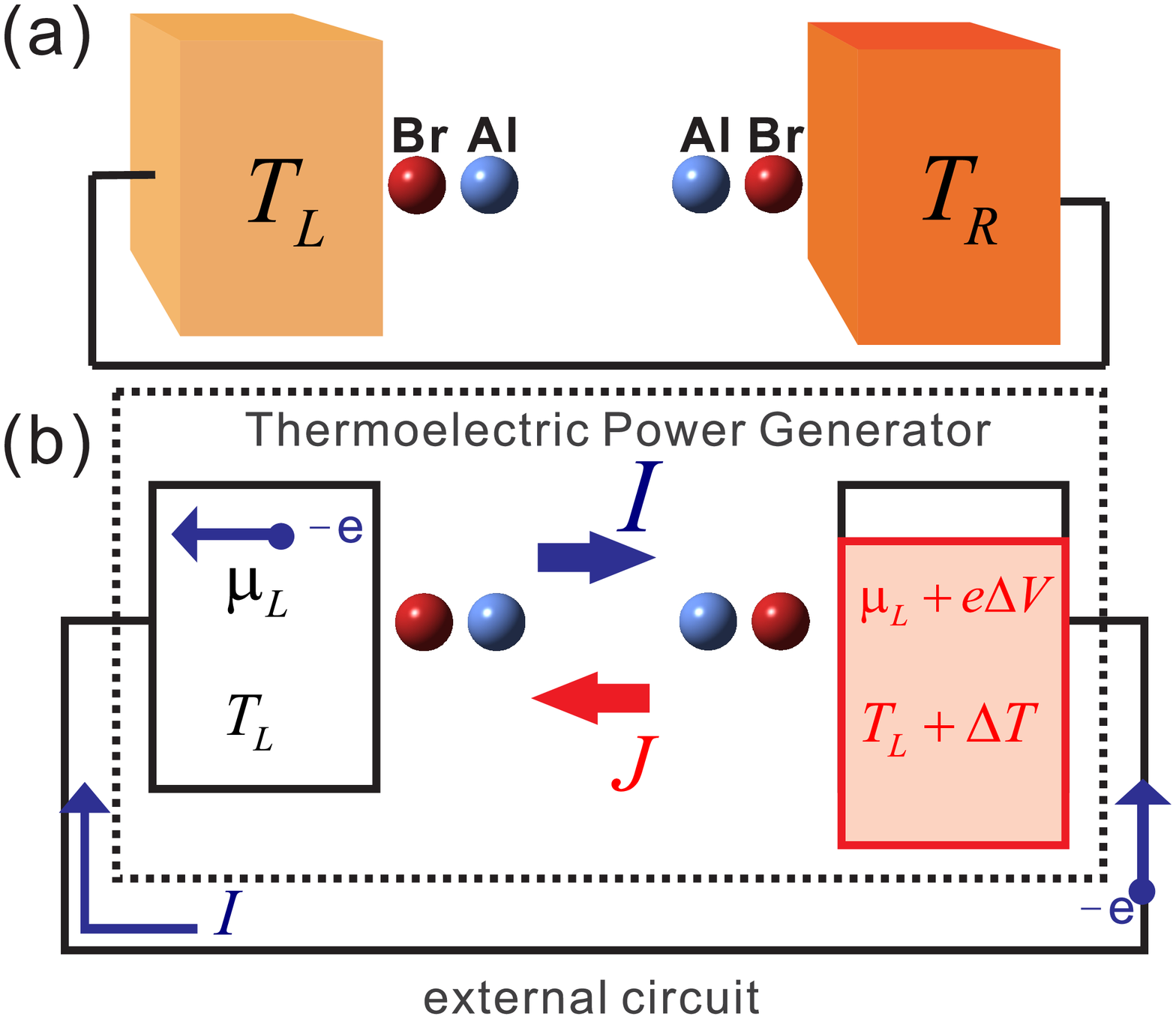}
\caption{
(color online)
(a) Schematics of the thermoelectric junction. The Br-jellium, Al-Br, and
Al-Al distance are $1.8$, $4.9$, and $8.6$~a.u., respectively. The Al-Al
distance has been significantly enlarged.
(b) The directions of electric current $I$ and thermal current
$J$ for $S<0$. Inside the power generator, current
travels from the lower to the higher potential, which is opposite of what
occurs when the nanojunction is a passive element in a circuit.}%
\label{Fig1}%
\end{figure}

Thermoelectric power generators in bulk systems employ electron gas as a
working fluid. It directly converts thermal energy into electric energy using
the Seebeck effect. Provided a temperature difference is maintained across the
device, it can generate electric power converted from the thermal energy. In
this paper, we explore the energy conversion mechanism of nanojunctions. We
present a parameter-free first-principles calculations for a thermoelectric
power generator in a truly atomic-scale system. To gain further insight into
the mechanism of energy conversion, we also developed an analytical theory for
it. As a specific example, we consider an atomic junction depicted in
Fig.~\ref{Fig1}(a), where the left and right electrodes serve as independent
cold- and hot-temperature reservoirs, respectively. This is not just an
academic example since recent studies have demonstrated the capability to
assemble one magnetic atom on a thin layer of atoms on the surface of
STM~\cite{Heinrich}. A similar technique might be applicable to a single Al
atom adsorbed onto a layer of Br atoms on the metal surface. In this regard,
the atomic junction could be formed by bringing two identical pieces of
metal-Br-Al surfaces close together before the reconstruction of Br and Al
atoms occurs.

As it was found in Ref.~\citenum{Lang2}, the paired metal-Br-Al junction
shows interesting device properties, such as negative differential resistance,
that utilize the relatively narrow density of states (DOSs) near the chemical
potentials. The narrow DOSs are due to the weak coupling between the Al atoms
and the electrodes via the \textquotedblleft spacer\textquotedblright~Br
atoms~\cite{Lyo, Lang3}. This junction also serves as an efficient
field-effect transistor because the narrow DOSs near chemical potentials can
be easily shifted by the gate voltages. As the Seebeck coefficients are
relevant to the slope of DOSs, the sharp DOSs result in a large magnitude in
the Seebeck coefficient~\cite{Liu}. When a temperature difference is
maintained between electrodes in a closed circuit, the Seebeck effect
generates an electromotive force ($emf$), which drives a current flowing
through the junction, hence the nanojunction can also be considered as an
efficient thermoelectric power generator~\cite{Justin}.

We investigate further the possibility of powering an atomic-scale device as a
field-effect transistor using heat instead of electricity. To illustrate this
point, we consider the paired metal-Br-Al junction in a three-terminal
geometry. When a finite temperature difference is maintained between
electrodes, the Seebeck effect converts the electron's thermal current into the
electric current flowing through the nanojunction. We observe that the
current's magnitude, polarity, and power on-off are controllable by the gate
field. Such current-voltage characteristics could be useful in the design of
nanoscale electronic devices, such as a transistor or switch.The nanojunction,
therefore, can be considered a transistor that employs the Seebeck effect to
power itself by converting the thermal energy into electric energy. The results
of this study may be of interest to researchers attempting to develop new
styles of thermoelectric nano-devices.

The flow of the discussion in this paper is as follows. In Sec.~II, we
describe density functional theory, theory of thermoelectricity, and theory of
thermoelectric power generator as a self-powered electronic device. In
Sec.~III, we discuss the thermoelectric properties of nanojunctions and the
thermoelectric power generator as a self-powered device. Finally, we conclude
our findings in Sec.~IV.

\section{Theoretical Methods}

We briefly present an introduction of the Lippmann-Schwinger (LS) equation and
density-functional theory (DFT) in subsection A. In subsection B, we present
the theory of the Seebeck coefficient, electric conductance and thermal
conductance. In subsection C, we present the theory to calculate the
Seebeck-induced electric current, voltage, power, and efficiency of energy
conversion. These quantities are calculated in the truly atomic scale junction
in terms of the nonequilibrium scattering wave-functions obtained
self-consistently in the framework of DFT+LS calculations. Both DFT+LS and
DFT+Keldysh nonequilibrium Green's function (NEGF) have been applied
extensively to a wide range of problems of nonequilibrium quantum transport of
device physics problems under finite biases from first-principles approaches.
The formal connection between DFT+LS and DFT+NEGF can be found in Ref.~\citenum{JianWang}.

\subsection{Density Functional Theory and Lippmann-Schwinger Equation}

We model a nanoscale junction as a passive element formed by two semi-infinite
metal with planar surfaces held a fixed distance apart connecting to an
external \ battery of bias $V_{B}$, with a nano-structured object bridging the
gap between them. The full Hamiltonian of the system is $H=H_{0}+V$, where
$H_{0}$ is the Hamiltonian due to the biased bimetallic electrodes that we
model as ideal metal (jellium model), and $V$ is the scattering potential of a
group of atoms bridging the gap. We assume that the left electrode is
positively biased such that $V_{B}=(\mu_{R}-\mu_{L})/e$, where $\mu_{L}=\mu$
and $\mu_{R}=\mu+eV_{B}$ are the chemical potential deep in the left and right
electrodes, respectively.

The unperturbed wave functions of the biased bimetallic electrodes have the
form, $\Psi_{E\mathbf{K}_{\parallel}}^{0,L(R)}(\mathbf{r})=e^{i\mathbf{K}%
_{\parallel}\mathbf{\cdot r}_{\bot}}\cdot u_{E\mathbf{K}_{\parallel}}%
^{L(R)}(z)$, where $\mathbf{r}_{\bot}$ is the coordinate parallel to the
surfaces and $z$ is the coordinate normal to them. Electrons are free to move
in the plane perpendicular to the $z$ direction, and $\mathbf{K}_{\parallel}$
is the momentum of electron in the plane parallel to the electrode surfaces.
Partial charges spilled from the electrode positive-background edges into the
vacuum region. This causes an electrostatic potential barrier between two
electrodes. The charge density distribution and effective single-particle wave
functions $u_{E\mathbf{K}_{\parallel}}^{L(R)}(z)$ are calculated in the
framework of the density functional formalism by solving the coupled
Shr\"{o}dinger and Poisson equations iteratively until self-consistency is
obtained~\cite{Lang4}. The wave function describes the electrons incident from
the right electrode, satisfying the following boundary condition :
\begin{equation}
u_{E\mathbf{K}_{\parallel}}^{R}(z)=(2\pi)^{-\frac{3}{2}}\times%
\genfrac{\{}{.}{0pt}{}{\frac{1}{\sqrt{k_{R}}}(e^{-ik_{R}z}+R\text{ }%
e^{ik_{R}z}),\text{ }z\rightarrow\infty,}{\frac{1}{\sqrt{k_{L}}}T\text{
}e^{-ik_{L}z},\text{ }z\rightarrow-\infty.}
\label{uk}%
\end{equation}

The group of atoms is considered in the scattering approaches. Corresponding
to each of the unperturbed wave functions, a Lippmann-Schwinger equation
involving a Green's function for the biased bimetallic junction is solved in
the plane wave basis, where a basis of approximately $2300$ plane waves has
been chosen for this study~\cite{Lang}:
\begin{widetext}%
\begin{equation}
\Psi_{E\mathbf{K}_{\parallel}}^{L(R)}(\mathbf{r})=\Psi_{E\mathbf{K}%
_{\parallel}}^{0,L(R)}(\mathbf{r})+\int d^{3}\mathbf{r}_{1}\int d^{3}%
\mathbf{r}_{2}G_{E}^{0}(\mathbf{r},\mathbf{r}_{1})V(\mathbf{r}_{1}%
,\mathbf{r}_{2})\Psi_{E\mathbf{K}_{\parallel}}^{L(R)}(\mathbf{r}%
_{2}),\label{Lippmann}%
\end{equation}
\end{widetext}
where $\Psi_{E\mathbf{K}_{\parallel}}^{L(R)}(\mathbf{r})$ represents the
nonequilibrium scattering wave function of the electrons with energy $E$
incident from the left (right) electrode. The quantity $G_{E}^{0}$ is the
Green's function for the biased bimetallic electrodes and $V(\mathbf{r}%
_{1},\mathbf{r}_{2})$ is the scattering potential electrons experience:
\begin{widetext}%
\begin{equation}
V(\mathbf{r}_{1},\mathbf{r}_{2})=V_{ps}(\mathbf{r}_{1},\mathbf{r}%
_{2})+\left\{  \left(  V_{xc}\left[  n\left(  \mathbf{r}_{1}\right)  \right]
-V_{xc}\left[  n_{0}\left(  \mathbf{r}_{1}\right)  \right]  \right)  +\int
d\mathbf{r}_{3}\frac{\delta n\left(  \mathbf{r}_{3}\right)  }{\left\vert
\mathbf{r}_{1}-\mathbf{r}_{3}\right\vert }\right\}  \delta(\mathbf{r}%
_{1}-\mathbf{r}_{2}),\label{V}%
\end{equation}
\end{widetext}
where $V_{ps}(\mathbf{r}_{1},\mathbf{r}_{2})$ is the electron-ion interaction
potential represented with pseudopotential; $V_{xc}\left[  n\left(
\mathbf{r}\right)  \right]  $ is the exchange-correlation potential calculated
at the level of the local-density approximation; $n_{0}\left(  \mathbf{r}%
\right)  $ is the electron density for the pair of biased bare electrodes;
$n\left(  \mathbf{r}\right)  $ is the electron density for the total system;
and $\delta n\left(  \mathbf{r}\right)  $ is their difference. The wave
functions of the entire system are calculated iteratively until
self-consistency is obtained~\cite{Lang}.

These right- and left-moving wave functions, weighed with the Fermi-Dirac
distribution function according to their energies and temperatures, are
applied to calculate the electric current as:
\begin{widetext}%
\begin{equation}
I(\mu_{L},T_{L};\mu_{R},T_{R})=\frac{e\hbar}{mi}\int dE\int
d\mathbf{r}_{\mathbf{\perp}}\int d\mathbf{K}_{\parallel}\left[  f_{E}(\mu
_{R},T_{R})I_{E{E},\mathrm{\mathbf{K}}_{||}}^{RR}(\mathbf{r})-f_{E}(\mu
_{L},T_{L})I_{E{E},\mathrm{\mathbf{K}}_{||}}^{LL}(\mathbf{r})\right]  ,
\label{I}%
\end{equation}
\end{widetext}
where $I_{E{E}^{\prime},\mathrm{\mathbf{K}}_{||}}^{RR(LL)}(\mathbf{r})=\left[
\Psi_{E,\mathrm{\mathbf{K}}_{||}}^{R(L)}(\mathbf{r})\right]  ^{\ast}\nabla
\Psi_{{E}^{\prime},\mathrm{\mathbf{K}}_{||}}^{R(L)}(\mathbf{r})-\nabla\left[
\Psi_{E,\mathrm{\mathbf{K}}_{||}}^{R(L)}(\mathbf{r})\right]  ^{\ast}\Psi
_{{E}^{\prime},\mathrm{\mathbf{K}}_{||}}^{R(L)}(\mathbf{r})$ and
$d\mathbf{r}_{\mathbf{\perp}}$ represents an element of the electrode surface.
Here, we assume that the left and right electrodes are independent electron
reservoirs, with the electron population described by the Fermi-Dirac
distribution function, $f_{E}(\mu_{L(R)},T_{L(R)})=1/\{\exp[\left(
E-\mu_{L(R)}\right)  /(k_{B}T_{L(R)})]+1\}$, where $\mu_{L(R)}$ and $T_{L(R)}$
are the chemical potential and the temperature in the left (right) electrode,
respectively. More detailed descriptions of theory can be found in Refs.~\citenum{Lang,DiVentra2,Chen}.

The above expression can be cast in a Landauer-B\"{u}ttiker formalism:%
\begin{equation}
I(\mu_{L},T_{L};\mu_{R},T_{R})=\frac{2e}{h}\int dE\tau(E)[f_{E}(\mu_{R}%
,T_{R})-f_{E}(\mu_{L},T_{L})], \label{Landau-Buttiker}%
\end{equation}
where $\tau(E)=\tau^{R}(E)=\tau^{L}(E)$ is a direct consequence of the
time-reversal symmetry, and $\tau^{R(L)}(E)$ is the transmission function of
the electron with energy $E$ incident from the right (left) electrode,
\begin{equation}
\tau^{R(L)}(E)=\frac{\pi\hbar^{2}}{mi}\int{d\mathbf{r}}_{\perp}{\int
{d\mathrm{\mathbf{K}}_{||}}}I_{EE,\mathrm{\mathbf{K}}_{||}}^{RR(LL)}%
(\mathrm{\mathbf{r}}). \label{tau}%
\end{equation}
Note that $e$ is positive in the definition of Eq.~(\ref{Landau-Buttiker}).

The electron's thermal current, defined as the rate at which thermal energy
flows from the right (into the left) electrode, is
\begin{widetext}%
\begin{equation}
J_{el}^{R(L)}(\mu_{L},T_{L};\mu_{R},T_{R})=\frac{2}{h}\int dE(E-\mu
_{R(L)})\tau(E)[f_{E}(\mu_{R},T_{R})-f_{E}(\mu_{L},T_{L})], \label{JJ}%
\end{equation}
\end{widetext}
respectively.

\subsection{Theory of the Thermoelectric Properties in the Nanoscale
Junctions}

We assume that the nanojunction is not connected to external circuit such that
$\mu_{R}=\mu_{L}=\mu$ and $T_{R}=T_{L}=T$, and then we consider an
infinitesimal current [$(dI)_{T}=I(\mu,T;\mu,T+dT)]$ induce by an
infinitesimal temperature difference $dT$ raised in the right electrode [i.e.,
$T_{R}=T_{L}+dT$]. The Seebeck effect generates a voltage difference $dV$ in
the right electrode [i.e., $\mu_{R}=\mu_{L}+dV$] which drives current
[$(dI)_{V}=I(\mu,T;\mu+edV,T)$]. The current cannot actually flow, thus,
$(dI)_{T}$ counterbalances $(dI)_{V}$ [i.e., $dI=(dI)_{T}+(dI)_{V}=0$]. The
Seebeck coefficient (defined as $S=\frac{dV}{dT}$) can be obtained by
expanding the Fermi-Dirac distribution functions in $(dI)_{T}$ and $(dI)_{V}$
to the first order in $dT$ and $dV$:%
\begin{equation}
S=-\frac{1}{eT}\frac{K_{1}(\mu,T)}{K_{0}(\mu,T)},\label{S}%
\end{equation}
where%
\begin{equation}
K_{n}(\mu,T)=-\int dE\tau(E)\left(  E-\mu\right)  ^{n}\frac{\partial f_{E}%
(\mu,T)}{\partial E}.\label{KKn}%
\end{equation}
The Seebeck coefficient up to the lowest order in temperatures is,%
\begin{equation}
S\approx\alpha T,\label{SlowT}%
\end{equation}
where $\alpha=-\pi^{2}k_{B}^{2}\frac{\partial\tau(\mu)}{\partial E}/\left(
3e\tau(\mu)\right)  $ \cite{Liu2}. Here, we have expanded $K_{n}(\mu,T)$ to
the lowest order in temperatures by using Sommerfeld expansion: $K_{0}%
\approx\tau(\mu)$, $K_{1}\approx\lbrack\pi^{2}k_{B}^{2}\tau^{\prime}%
(\mu)/3]T^{2}$, and $K_{2}\approx\lbrack\pi^{2}k_{B}^{2}\tau(\mu)/3]T^{2}$.
The Seebeck coefficient is positive (negative) when the slope of transmission
function is negative (positive) near the chemical potential.

When a finite temperature difference $\Delta T=T_{R}-T_{L}>0$ is raised in the
right electrode, the Seebeck effect generates a finite electromotive force,
$emf=\left\vert  \Delta V(T_{L},T_{R})\right\vert $, raising the potential
energy of the charge. We assume that the voltage difference $\Delta
V(T_{L},T_{R})$ is applied to the right electrode such that $\mu=\mu_{L}$ and
$\mu_{R}=\mu_{L} + e \Delta V(T_{L},T_{R})$. The voltage difference
$\Delta V(T_{L},T_{R})$ generated by the temperature difference $\Delta T$ is
approximately given by,
\begin{equation}
\Delta V(T_{L},T_{R})\approx\int_{T_{L}}^{T_{R}}S(\mu_{L}
,T)dT,\label{DeltaVSeebeck1}%
\end{equation}
where the Seebeck coefficient $S(\mu,T)$ is given by Eq.~(\ref{S}) with $\mu=\mu_{L}$.
Equation~(\ref{DeltaVSeebeck1}) becomes exact when $\Delta
T\rightarrow0$. The sign of $\Delta V(T_{L},T_{R})$ depends on the sign of the
Seebeck coefficient $S(\mu,T)$. For example, $\Delta V(T_{L},T_{R})<0$ for
$S<0$, as shown in Fig.~\ref{Fig1}(b).

Equation~(\ref{DeltaVSeebeck1}) is an approximation when $\Delta T$ is finite.
We have to remind ourselves of the fact that the left chemical potential
$\mu_{L}$ differs from the right chemical potential $\mu_{R}$ when the
temperature difference $\Delta T$ is large. In this case, the Seebeck
coefficient becomes relevant to both $\mu_{L}$ and $\mu_{R}$ with a more
complicated form~\cite{Liu},%

\begin{equation}
S(\mu_{L},T_{L};\mu_{R},T_{R})=-\frac{1}{e}\frac{K_{1}(\mu_{L},T_{L}%
)/T_{L}+K_{1}(\mu_{R},T_{R})/T_{R}}{K_{0}(\mu_{L},T_{L})+K_{0}(\mu_{R},T_{R}%
)},\label{Seebeck_finiteT}%
\end{equation}
where $K_{0}(\mu_{L(R)},T_{L(R)})$ and $K_{1}(\mu_{L(R)},T_{L(R)})$ are given
by Eq.~(\ref{KKn}). A more elaborate evaluation of the voltage difference
$\Delta V(T_{L},T_{R})$ can be performed using,%
\begin{widetext}
\begin{equation}
\Delta V(T_{L},T_{R})\approx\int_{0}^{\Delta T/2}S(\mu-\frac{e\Delta
V(\overline{T}-T,\overline{T}+T)}{2},\overline{T}-T;\mu+\frac{e\Delta
V(\overline{T}-T,\overline{T}+T)}{2},\overline{T}+T)dT,\label{DeltaV_exact}%
\end{equation}
\end{widetext}
where $\overline{T}=(T_{L}+T_{R})/2$, $\Delta T=T_{R}-T_{L}$ and $\mu=E_{F}$.

The induced voltage by the Seebeck effect using Eq.~(\ref{DeltaVSeebeck1}) and
(\ref{DeltaV_exact}) have been numerically verified, showing that Eq.~(\ref{DeltaVSeebeck1})
is a good approximation when the transmission function
$\tau(E)$ does not change rapidly around the chemical potentials and when
$\Delta T$ is not too large. Accordingly, we evaluate $\Delta V(T_{L},T_{R})$
throughout this study using Eq.~(\ref{DeltaVSeebeck1}) instead of Eq.~(\ref{DeltaV_exact})
for simplicity.

Corresponding to the electric current, the extra electron's thermal current
induced by an infinitesimal temperature ($dT$) and voltage ($dV$) across the
junctions is,
\begin{equation}
dJ_{el}=(dJ_{el})_{T}+(dJ_{el})_{V}. \label{extrJel1}%
\end{equation}
where $(dJ_{el})_{T}=J_{el}(\mu,T;\mu,T+dT)$ and $(dJ_{el})_{V}=J_{el}%
(\mu,T;\mu+edV,T)$ can be calculated by Eq.~(\ref{JJ}).

The electron's thermal conductance (defined as $k_{el}=dJ_{el}/dT$) can be
decomposed into two components:
\begin{equation}
\kappa_{el}(\mu,T)=\kappa_{el}^{T}(\mu,T)+\kappa_{el}^{V}(\mu,T),
\label{thercond}%
\end{equation}
where $\kappa_{el}^{T}=(dJ_{el})_{T}/dT$ and $\kappa_{el}^{V}=(dJ_{el}%
)_{V}/dT$. Using Sommerfeld expansion, they can be can be written as,%
\begin{equation}
\kappa_{el}^{T}(\mu,T)=\frac{2}{h}\frac{K_{2}(\mu,T)}{T}, \label{kelT0}%
\end{equation}
and%
\begin{equation}
\kappa_{el}^{V}(\mu,T)=\frac{2e}{h}K_{1}(\mu,T)S(\mu,T), \label{kelV0}%
\end{equation}
where $K_{1}(\mu,T)$ and $K_{2}(\mu,T)$ are given by Eq.~(\ref{KKn}). One
should note that $\kappa_{el}^{V}=0$ if the Seebeck coefficient of the system
is zero.

By expanding $K_{n}(\mu,T)$ and $S(\mu,T)$ to the lowest order in
temperatures, Eqs.~(\ref{kelT0}) and (\ref{kelV0}) become
\begin{equation}
\kappa_{el}^{V}\approx\beta_{V}T\text{ }^{3}\text{and }\kappa_{el}^{T}%
\approx\beta_{T}T, \label{kelTT}%
\end{equation}
where $\beta_{V}=-2\pi^{4}k_{B}^{4}[\tau^{\prime}(\mu)]^{2}/[9h\tau(\mu)]$ and
$\beta_{T}=2\pi^{2}k_{B}^{2}\tau(\mu)/(3h)$. In the above expansions, we have
applied the Sommerfeld expansions: $K_{1}(\mu,T)\approx\lbrack\pi^{2}k_{B}%
^{2}\tau^{\prime}(\mu)/3]T^{2}$, $K_{2}(\mu,T)\approx\lbrack\pi^{2}k_{B}%
^{2}\tau(\mu)/3]T^{2}$, and Eq.~(\ref{SlowT}).

In the limit of zero bias, Eq.~(\ref{Landau-Buttiker}) yields the electric
conductance (defined as $\sigma(T)=dI/dV$ ),
\begin{equation}
\sigma(T)=\frac{2e^{2}}{h}\int dEf_{E}\left(  \mu,T\right)  [1-f_{E}\left(
\mu,T\right)  ]\tau(E)/(k_{B}T), \label{cond}%
\end{equation}
which is relatively insensitive to temperatures if tunneling is the major
transport mechanism.

\begin{figure}[ptb]
\includegraphics[width=7cm]{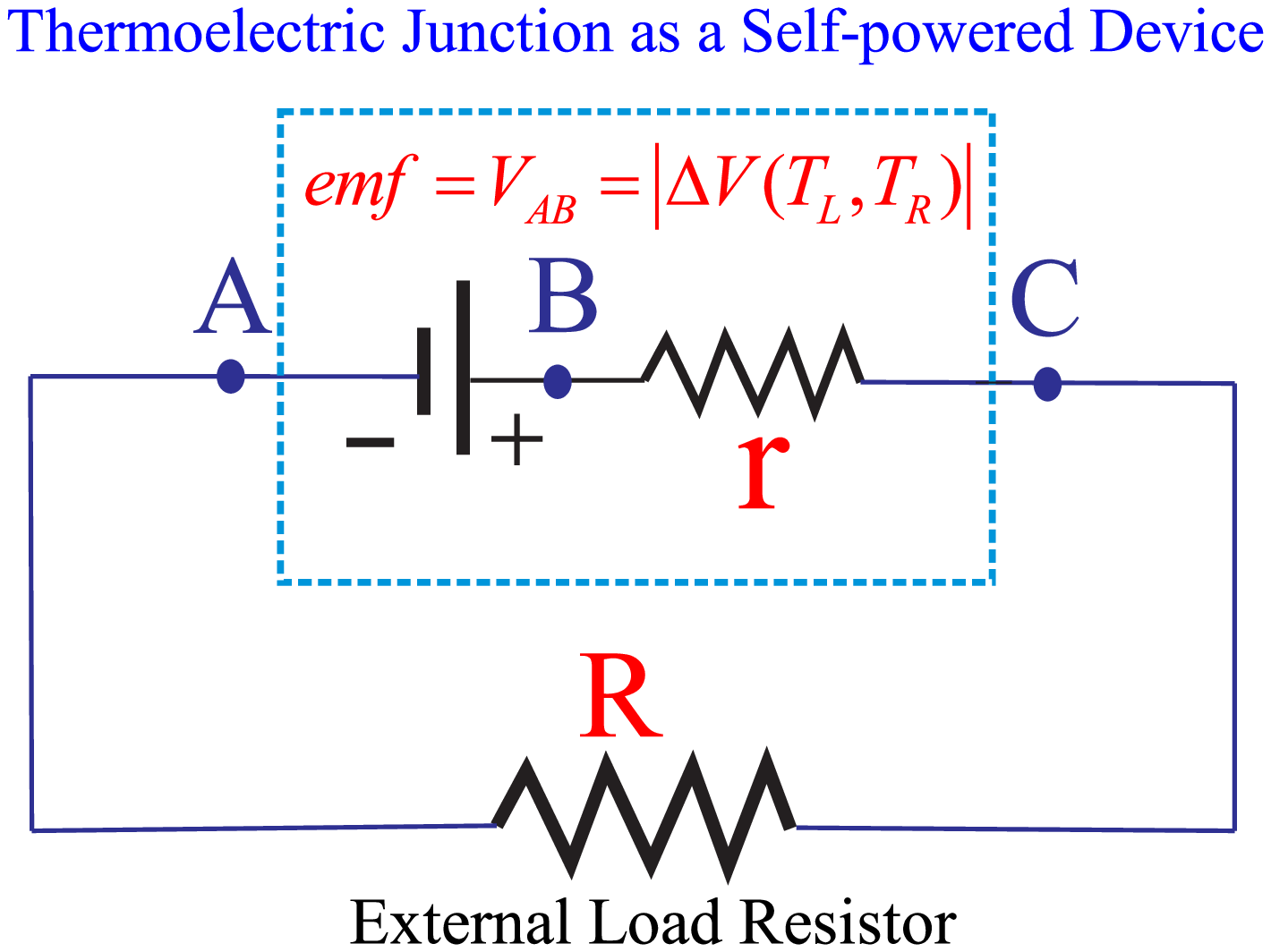}
\caption{
(color online)
The circuit diagram of the thermoelectric power generator depicted in
Fig.~\ref{Fig1}(b) as a self-powered electronic device.
The thermoelectric nanojunction can be represented as an ideal battery of
$emf=V_{AB}$ generated by the Seebeck effect and an internal
resistance $r$ for the nanojunction as an electronic device. We assume
that the external resistor $R\rightarrow0$ in this study for simplicity. }%
\label{Fig2}%
\end{figure}

\subsection{Theory of the Thermoelectric Power Generator as a Self-powered
Electronic Device}

In this subsection, we present a closed-circuit theory of the thermoelectric
nanojunction as a power generator and a self-powered electronic device by itself.
We consider the nanojunction with the Seebeck coefficient $S<0$
and $\Delta T=T_{R}-T_{L}>0$ such that the open-circuit Seebeck-effect induced
$emf$ is $|\Delta V(T_{L},T_{R})|$, as depicted in Fig.~\ref{Fig1}(b).
Figure~\ref{Fig1}(b) can be represented as a circuit
diagram shown in Fig.~\ref{Fig2}, where the thermoelectric junction is
considered to consist an ideal battery of $emf$ plus an internal resistance $r$.
The thermoelectric junction employs the Seebeck effect to generate an $emf$
and converts thermal energy into electric energy. Concurrently, the nanojunction
itself serves as a passive element with resistance $r$ which consumes electric energy.

We choose clockwise as positive such that the induced current $I>0$
for $S<0$. Beginning at point $A$ as the current traverses the circuit in the
positive direction, we obtain from Kirchhoff's loop rule,
\begin{equation}
V_{AB}-I r-I R=0,\label{Kirchhoff}%
\end{equation}
where the open-circuit $emf$ is $V_{AB}=-\Delta V(T_{L},T_{R})>0$ for $S<0$,
$r$ is the internal resistance of the nanojunction, and $R$ is the resistance of the
external resistor, as depicted in Fig.~\ref{Fig2}. Note that the current $I$ encounters a potential
increase due to the source of $emf$ between points $A$ and $B$, a potential drop [$V_{BC}=I r$]
due to the internal resistance between points $B$ and $C$, and a potential drop [$V_{CA}=I R$]
as the current traverses the external resistor with resistance $R$
between points $C$ and $A$, respectively.

In particular, we carry out this research considering $R \rightarrow 0$ such
that the voltage drop between points $C$ and $A$ is $V_{CA}=I R=0$. It yields%
\begin{equation}
V_{AB}= I r. \label{Kirchhoff2}%
\end{equation}
It implies that external circuit does not consume electric energy because the potential
difference across two terminals of the
thermoelectric power generators (i.e., the terminal voltage) is zero due to
$R \rightarrow 0$. This simplified scenario helps us to analyze the detailed
mechanism of energy conversion between the thermal current and electric current.
Such analysis will provide us an insight into the device physics of the
thermoelectric junctions as self-powered electronic devices.

The current $I=\sigma V_{AB}$ can be evaluated by Eq.~(\ref{Kirchhoff2}), where
$V_{AB}$ is the Seebeck-effect induced $emf=|\Delta V_{T_{L},T_{R}}|$
[which is evaluated by Eq.~(\ref{Seebeck_finiteT})] and $\sigma \approx 1/r$ is the
conductance of nanojunction [which is evaluated by Eq.~(\ref{cond})]. We have numerically
verified and confirmed that
$I=(I)_{\Delta V}=(I)_{\Delta T}$, where%
\begin{widetext}
\begin{equation}
(I)_{\Delta V}=-\frac{2e}{h}\int dE\tau(E)[f_{E}(\mu_{L}+e\Delta
V(T_{L},T_{R}),T_{L})-f_{E}(\mu_{L},T_{L})], \label{Ifinite2}%
\end{equation}
\end{widetext}
and%
\begin{equation}
(I)_{\Delta T}=\frac{2e}{h}\int dE\tau(E)[f_{E}(\mu_{L},T_{L}+\Delta
T)-f_{E}(\mu_{L},T_{L})].\label{Ifinite3}%
\end{equation}

The above two equations have the sign convention conformed to the direction of
current depicted Fig.~\ref{Fig1}(b) and Fig.~\ref{Fig3} for $S<0$. We may
interpret $(I)_{\Delta T}$ as the current which traverses the
thermoelectric nanojunction as a "pure" source of $emf$ generated by the
temperature difference $\Delta T$ using the Seebeck effect. Similarly, $(I)_{\Delta V}$
is the current which traverses the nanojunction as an internal
resistor. Equations~(\ref{Ifinite2}) and (\ref{Ifinite3})
describe the flow of electrons associated with the probability flux.

The flow of probability density can also transport energy as electrons travel between
two electrodes. The electrons that travel with energy $E$ from the right (left) electrode
carry an amount of energy $(E-\mu_{R(L)})$. Correspondingly, the $\Delta V$-induced
energy current carried by the probability current flowing out of the right electrode (into to
the left electrode) is,
\begin{widetext}
\begin{equation}
(J_{el}^{R(L)})_{\Delta V}=-\frac{2}{h}\int dE(E-\mu_{R(L)}%
)\tau(E)[f_{E}(\mu_{L}+e\Delta V(T_{L},T_{R}),T_{L})-f_{E}(\mu_{L}%
,T_{L})],\label{JRLfiniteV}%
\end{equation}
\end{widetext}
where $\mu_{R}=\mu_{L}+e\Delta V(T_{L},T_{R})$ and $\Delta V(T_{L},T_{R})$ is
given by Eq.~(\ref{DeltaVSeebeck1}). Similarly, the $\Delta T$-induced electron's thermal
current carried by the probability current flowing out of the right electrode
(into to the left electrode) is,
\begin{equation}
(J_{el})_{\Delta T}=\frac{2}{h}\int dE(E-\mu_{L})\tau(E)[f_{E}%
(\mu_{L},T_{L}+\Delta T)-f_{E}(\mu_{L},T_{L})].\label{JRLfiniteT}%
\end{equation}
Note that $(J_{el}^{R})_{\Delta V}$ and $(J_{el}^{L})_{\Delta
V}$ are positive and $(J_{el}^{R})_{\Delta V}>(J_{el}%
^{L})_{\Delta V}$ for $S<0$. The sign convention defined in
Eqs.~(\ref{Ifinite2})-(\ref{JRLfiniteT}) comply with the direction depicted in
Fig.~\ref{Fig1}(b) and  Fig.~\ref{Fig2} for $S<0$.

Subtracting $(J_{el}^{L})_{\Delta V}$ from $(J_{el}^{R})_{\Delta
V}$, we obtain
\begin{equation}
\Delta P=(J_{el}^{R})_{\Delta V}-(J_{el}^{L})_{\Delta
V}>0,\label{Power}%
\end{equation}
where $\Delta P$ is the electric power delivered by the thermoelectric
junction served as a "pure" battery which employs the Seebeck effect to convert
the electron's thermal current into electric energy. If the external resistor
with resistance $R \rightarrow 0$, then the external circuit does not consume
electric energy. In this case, the nanojunction simultaneously
serves as an electronic device which consumes the entire electric power
generated by itself, %
\begin{equation}
\Delta P\approx\sigma(\Delta V)^{2}\approx\int_{T_{L}}^{T_{R}}SdT\int_{T_{L}%
}^{T_{R}}S\sigma dT.\label{Power2}%
\end{equation}
Note that Eq.~(\ref{Power2}) can alternatively be expressed as,%
\begin{equation}
\Delta P=-\Delta V(T_{L},T_{R}) \cdot I(T_{L},T_{R})>0.\label{Power3}%
\end{equation}
Eqs.~(\ref{Power})-(\ref{Power3}) have been numerically verified to be
consistent with each other. If the system is n-type (i.e., $S<0$), then
$\Delta V(T_{L},T_{R})<0$ and $I(T_{L},T_{R})>0$, as shown in
Fig.~\ref{Fig1}(b). Similarly, if the system is p-type (i.e., $S>0$), then
$\Delta V(T_{L},T_{R})>0$ and $I(T_{L},T_{R})<0$. Both cases render
$\Delta P>0$.

\section{Results and Discussion}

Provided that a finite temperature difference is raised between electrodes,
the Seebeck effect will generate an electric current. It is observed that the
nanojunction itself can be deemed a field-effect transistor, where gate field
can control the current. Thus and so, the atomic junction in three-terminal
geometry  can be considered as an electronic device which can be self-powered
by providing a temperature difference across the junction.

As an a specific example, we consider an atomic-scale junction in a
three-terminal geometry as a thermoelectric power generator, depicted in
Fig.~\ref{Fig6}(a). In the paired metal-Br-Al junction, we assume that the distance
between two Al atoms is sufficiently large. The interaction between two pieces
of metal-Br-Al is minimal, and in this fashion, no rearrangement of atom
positions occurs. Accordingly, the phonon's thermal current could be
suppressed because of poor mechanical coupling between two pieces of
metal-Br-Al is minimal. We neglect it for simplicity in this study.

\subsection{Thermoelectric Properties of the Junction}

In an open circuit, an infinitesimal temperature $dT$ raised in the right
electrode induces $(dI)_{T}$ and $(dI)_{V}$ which counterbalance each other.
This renders the Seebeck coefficient defined in Eq.~(\ref{S}), as described in
subsection IIB.

Correspondingly, the electron's thermal conductance is defined as $\kappa
_{el}=dJ_{el}/dT$, as given by Eq.~(\ref{thercond}). The electron's thermal
conductance can be decomposed into two components, $\kappa_{el}=\kappa
_{el}^{T}+\kappa_{el}^{V}$, where $\kappa_{el}^{T}$ and $\kappa_{el}^{V}$ are
given by Eqs.~(\ref{kelT0}) and (\ref{kelV0}), respectively.

\begin{figure}[ptb]
\includegraphics[width=8cm]{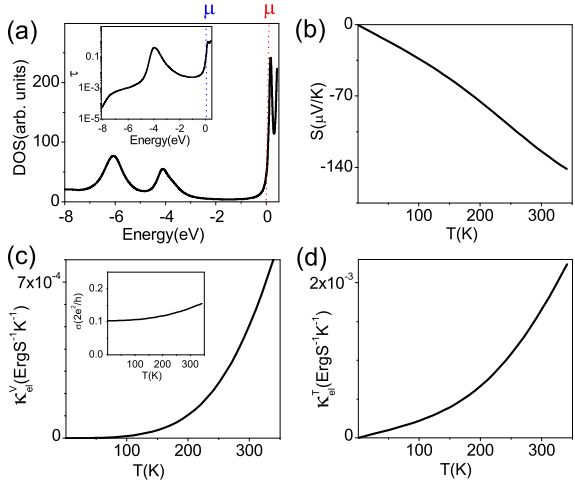}
\caption{
(color online)
(a) The DOSs (inset: transmission functions) as a function of energies;
(b) the Seebeck coefficient S; (c) the thermal conductance induced by the
temperature difference $(\kappa_{el}^{V})$ (inset: conductance $\sigma$);
and (d) the thermal conductance induced by the voltage difference
($\kappa_{el}^{T})$ as a function of temperatures.}%
\label{Fig3}%
\end{figure}

Fig.~\ref{Fig3}(a) shows that the system is characterized by a sharp
transmission function corresponding to a narrow DOSs near the chemical
potential. As it was found in Ref.~\citenum{Lang2}, the DOSs become narrower
when the distance between two Al atoms increases. The narrow DOSs result in
substantial Seebeck coefficients, as shown in Fig.~\ref{Fig3}(b). It shows
that the thermoelectric junction is n-type ($S<0$) since the narrow DOSs are
slightly above the chemical potential and the slope of the transmission
$\tau^{\prime}(\mu)$ is positive. At low temperatures (about $T<100$~K), the
Seebeck coefficients remain linear ($S\approx\alpha T$), as described in
Eq.~(\ref{SlowT}). We present the absolute values of $\kappa_{el}^{T}$ and
$\kappa_{el}^{V}$ as functions of temperatures in Fig.~\ref{Fig3}(c)~and~(d),
respectively. At low temperatures, $\kappa_{el}^{V}\approx-\beta_{V}T^{3}$ and
$\kappa_{el}^{T}\approx\beta_{T}T$, as shown in Eq.~(\ref{kelTT}). The inset
of Fig.~\ref{Fig3}(c) shows the zero-bias electric conductance as a function
of temperatures calculated from Eq.~(\ref{cond}).

\subsection{The Junction as a Thermoelectric Power Generator and a
Self-powered Electronic Device}

Suppose that the junction is not connected to battery (i.e., $\mu_{R}=\mu
_{L}=\mu$) and $T_{R}=T_{L}=T$. Provided that a finite temperature difference
$\Delta T=T_{R}-T_{L}$ is raised in the right electrode (i.e., $T_{L}=T$ and
$T_{R}=T+\Delta T$), the Seebeck effect generates a finite electromotive force
$emf=\left\vert \Delta V(T_{L},T_{R})\right\vert $, where $\Delta
V(T_{L},T_{R})$ is given by Eq.~(\ref{DeltaVSeebeck1}). This voltage
difference induces a net electric current $I(T_{L},T_{R})$ which
travels from lower to higher potential inside the power generator, as
described in Eq.~(\ref{Ifinite2}) and Fig.~\ref{Fig1}(b). Correspondingly, the
electron's thermal current $(J_{el}^{R})_{\Delta V}$
[Eq.~(\ref{JRLfiniteV})] absorbs energy in heat from the hot (right)
electrode, produces electric power $\Delta P$ [Eq.~(\ref{Power})] using the
Seebeck effect, and then gives up energy in heat to the cold (left) electrode
via $(J_{el}^{L})_{\Delta V}$ [Eq.~(\ref{JRLfiniteV})], as described in
Fig.~\ref{Fig1}(b) for $S<0$.

\begin{figure}[ptb]
\includegraphics[width=8cm]{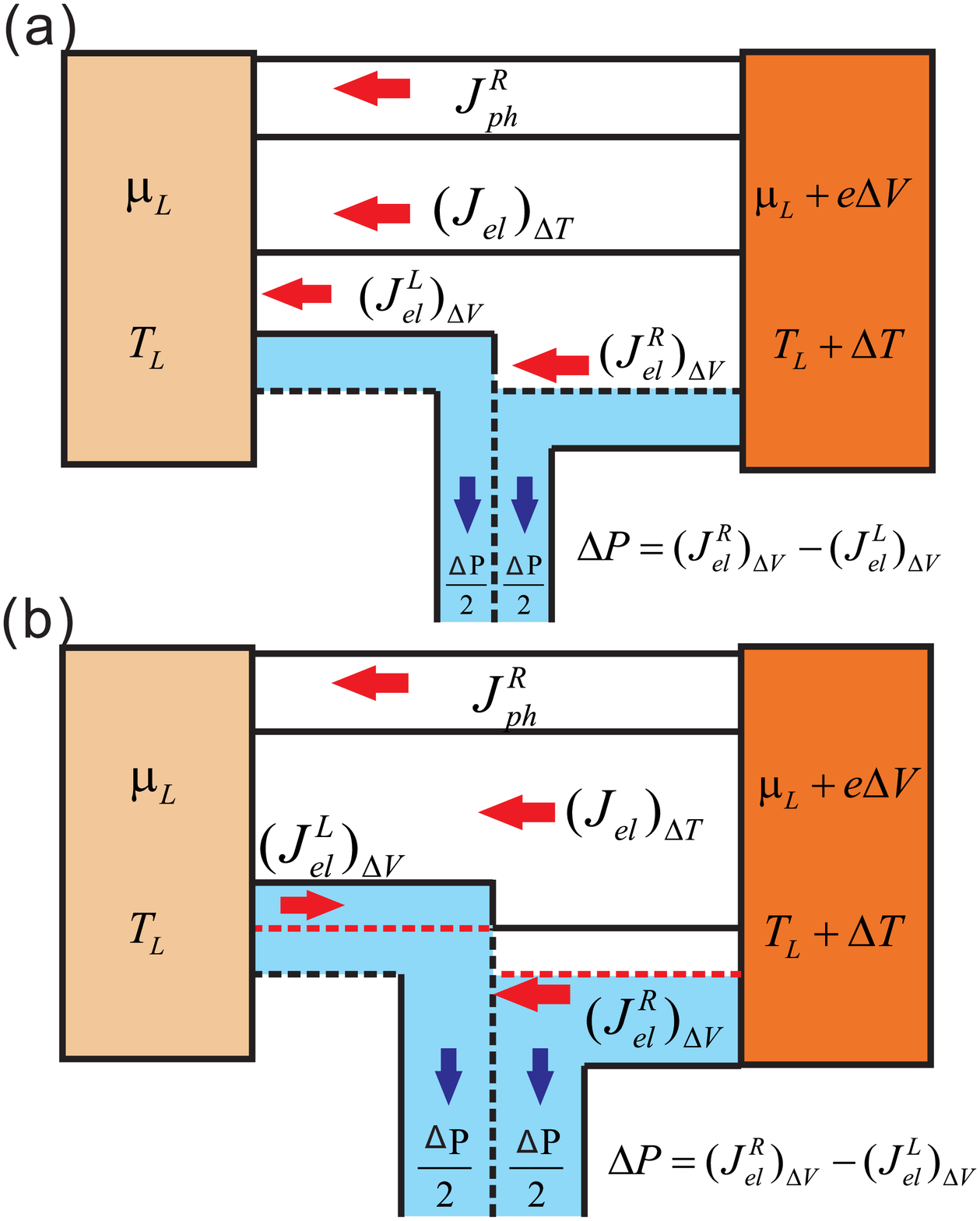}
\caption{
(color online)
(a) and (b) are the schematic representation of the
power generator for $(J_{el}^{L})_{\Delta V}>0$ and $(J_{el}%
^{L})_{\Delta V}<0$, respectively. Each electrode provides one half of the
electric power $\Delta P = (J_{el}^{R})_{\Delta V}-(J_{el}%
^{L})_{\Delta V}$ using the Seebeck effect. The thermal currents $
J_{ph}^{R}$ and $\left( J_{el}\right)  _{\Delta T}$ are driven by
the temperature difference $\Delta T$ and both are not converted into
electric energy.}%
\label{Fig4}
\end{figure}

To gain further insight into the mechanism of energy conversion, we
theoretically analyze the electron's thermal current. Applying Sommerfeld
expansion to $\left( J_{el}^{R}\right)  _{\Delta V}$ and $\left(
J_{el}^{L}\right)  _{\Delta V}$ in Eq.~(\ref{JRLfiniteV}), we obtain%
\begin{equation}
\left(J_{el}^{R}\right)  _{\Delta V}\approx\sigma(\Delta
V)^{2}T/\Delta T+\sigma(\Delta V)^{2}/2,\label{JRDV4}%
\end{equation}
and%
\begin{equation}
\left(J_{el}^{L}\right)  _{\Delta V}\approx\sigma(\Delta
V)^{2}T/\Delta T-\sigma(\Delta V)^{2}/2,\label{JLDV4}%
\end{equation}
where we have simplified Eqs.~(\ref{JRDV4}) and (\ref{JLDV4}) using the
following approximations: $\tau(\mu+e\Delta V)\approx$ $\tau(\mu)$,
$\tau^{\prime}(\mu+e\Delta V)\approx$ $\tau^{\prime}(\mu)$, $\sigma
\approx2e^{2}\tau(\mu)/h$, $S\approx-\pi^{2}k_{B}^{2}\frac{\partial\tau(\mu
)}{\partial E}T/\left(  3e\tau(\mu)\right)  $, and $\Delta V\approx S\Delta
T$.

Equations.~(\ref{JRDV4}) and (\ref{JLDV4}) immediately provides the law of
energy conservation as further described below. The thermoelectric junction
serves as an ideal source of $emf$ which delivers electric power $\Delta P=\left(J_{el}%
^{R}\right)  _{\Delta V}-\left(J_{el}^{L}\right)  _{\Delta V}$ converted from the
electron's thermal currents using the Seebeck effect. Concurrently, the nanojunction
itself serves as a passive element which consumes electric power dissipated by the
internal resistor at a rate of $\sigma (\Delta V^{2})$.

Equations~(\ref{JRDV4}) and (\ref{JLDV4}) imply that each electrode approximately
provides one half of the electric power using the Seebeck effect and the electric
power is mainly converted from $\left(J_{el}\right)  _{\Delta V}$. No
energy conversion is possible when the Seebeck coefficients is vanishing
because $\left(J_{el}^{R}\right)  _{\Delta V}=0$. We note that
$J_{el}^{R}=(J_{el}^{R})_{\Delta V}+(J_{el})_{\Delta
T}$ removes heat energy from the hot electrode, converts heat energy into
electric energy, and rejects waste heat to the cold electrode by $J_{el}^{L}$.
Both $(J_{el})_{\Delta T}$ and $J_{ph}^{R}$ do
not play active role in the energy conversion. The above discussions are
summarized in Fig.~\ref{Fig4}(a).

The energy current $\left(J_{el}^{L}\right)  _{\Delta V}$ flowing
into the cold electrode could be negative when $\Delta T$ is sufficiently
large, as shown in Fig.~\ref{Fig4}(b). To show this, we integrate
Eq.~(\ref{DeltaVSeebeck1}) using Eq.~(\ref{SlowT}), $K_{1}(\mu,T)\approx
\lbrack\pi^{2}k_{B}^{2}\tau^{\prime}(\mu)/3]T^{2}$, $K_{2}(\mu,T)\approx
\lbrack\pi^{2}k_{B}^{2}\tau(\mu)/3]T^{2}$. We obtain $\Delta V\approx\alpha
T^{2}[(\Delta T/T)^{2}+2(\Delta T/T)]/2$, where $\alpha=-[\pi^{2}k_{B}%
^{2}\frac{\partial\tau(\mu)}{\partial E}/\left(  3e\tau(\mu)\right)  ]$.
Together with Eq.~(\ref{JLDV4}), we arrive at%
\begin{equation}
\left(J_{el}^{L}\right)  _{\Delta V}\approx\sigma ST\Delta
V\{1-[(\Delta T/T)+(\Delta T/T)^{2}]/4\}, \label{JLDV5}%
\end{equation}
from which we observe that $\left(J_{el}^{L}\right)  _{\Delta V}$
changes sign at $\Delta T\approx1.236$~$T$ (which is universal), and $\left(
J_{el}^{L}\right)  _{\Delta V}<0$ when $\Delta T\gtrapprox1.236$~$T$.
Fig.~\ref{Fig4}(a) and \ref{Fig4}(b) illustrate the schematic representation
of energy conversion for positive and negative $\left(J_{el}%
^{L}\right)  _{\Delta V}$, respectively. Fig.~\ref{Fig4}(a) and \ref{Fig4}(b)
illustrate the schematic representation of energy conversion for positive and
negative $\left(J_{el}^{L}\right)  _{\Delta V}$, respectively.

Let us focus on the energy conversion efficiency $\eta_{el}$ defined as the
ratio of the electric power $\Delta P$ generated by the Seebeck effect to the
electron's thermal current $J_{el}^{R}=(J_{el}^{R})_{\Delta
V}+(J_{el})_{\Delta T}$ which removes thermal energy from the high
temperature reservoir:%
\begin{equation}
\eta_{el}=\frac{\Delta P}{J_{el}^{R}},\label{efficiency}%
\end{equation}
where $\Delta P$, $(J_{el}^{R})_{\Delta V}$, and  $(J_{el})_{\Delta T}$
are given by Eq.~(\ref{Power3}), (\ref{JRLfiniteV}), and
(\ref{JRLfiniteT}), respectively. In the above definition, we have neglect
other effects [\emph{e.g.} the phonon's thermal current $J_{ph}^{R}$
and possible photon radiation] which transfer energy from the hot to cold
electrodes. From this viewpoint, the energy conversion efficiency presented in
this study may be overestimated. We note that the size of the paired
metal-Br-Al system is small. This may lead to suppression of the photon
radiation because photon radiation is proportional the surface area. Moreover,
the paired metal-Br-Al system has a poor mechanical link between electrodes
when the distance of two Al atoms is sufficiently far apart. The poor
mechanical coupling between two phonon reservoir could lead to suppression of
the phonon's thermal current. These two features are helpful to increase the
energy conversion efficiency. Note that the phonon's thermal current and
photon radiation do not affect the magnitudes of the current and electric
power generated by the Seebeck effect.

\begin{figure}[ptb]
\includegraphics[width=8cm]{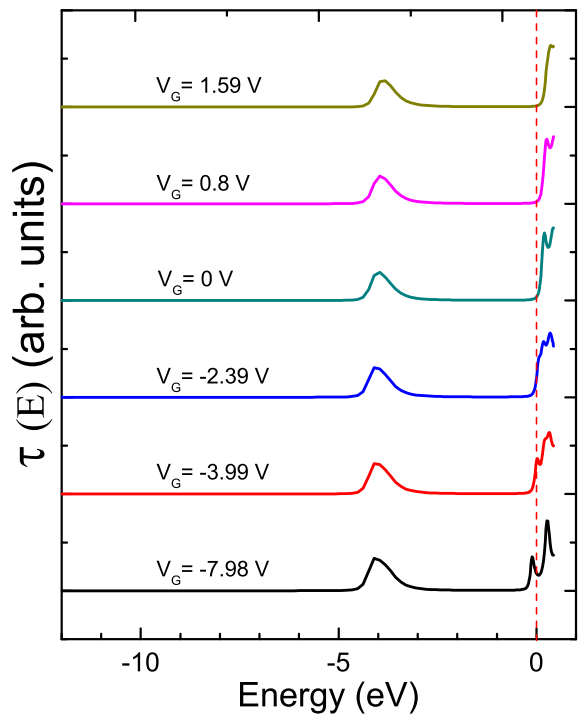}
\caption{
(color online)
Transmission functions as functions of energies for various gate voltages ($V_{G}=$ -7.98,
-3.99, -2.39, 0, 0.8, 1.59 V). The Fermi level is set to be zero of energy.}%
\label{Fig5}%
\end{figure}

To show that the atomic junction can be considered as a field-effect
transistor which can be self-powered, we consider the nanojunction in a
\textit{three-terminal} geometry with a finite temperature $\Delta T=100$ K
maintained between electrodes (where $T_{L}=200$ K and $T_{R}=300$ K). The
gate field is introduced as a capacitor composed of two parallel circular
charged disks separated at a certain distance from each other. The axis of the
capacitor is perpendicular to the transport direction. One plate is placed
close to the nano-object while the other plate, placed far away from the
nano-object, is set to be the zero reference
field~\cite{DiVentragate,Solomon1,Solomon2,Ma}.

\begin{figure}[ptb]
\includegraphics[width=8cm]{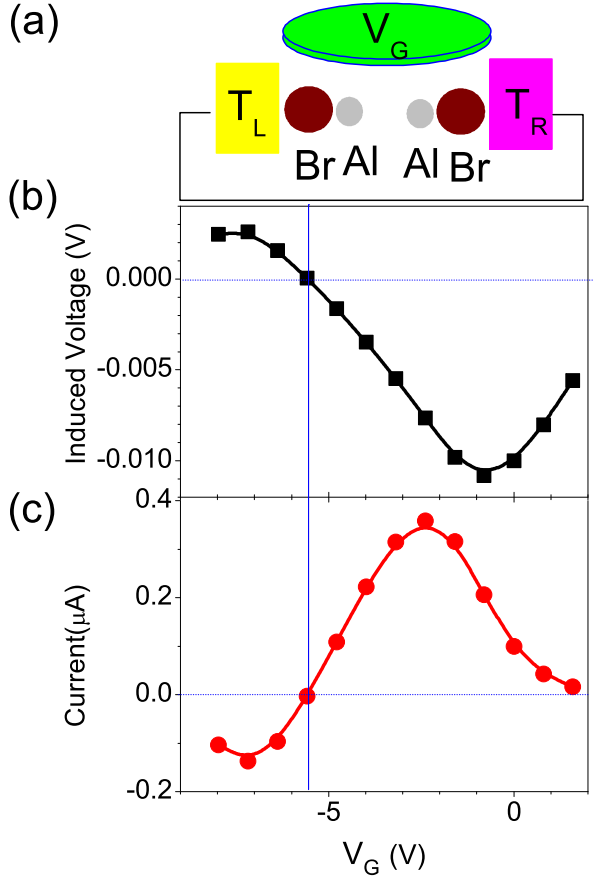}
\caption{
(color online)
(a) Schematic of the three-terminal junction. The Seebeck-effect induced
(b) voltage $V(T_{L},T_{R})$ and (c) current $I(T_{L},T_{R})$
as functions of gate voltages via the temperature difference
$\Delta T=T_{R}-T_{L}$, where $T_{L}= 200$~K and $T_{R}=300$~K.}%
\label{Fig6}%
\end{figure}
\begin{figure}[ptb]
\includegraphics[width=8cm]{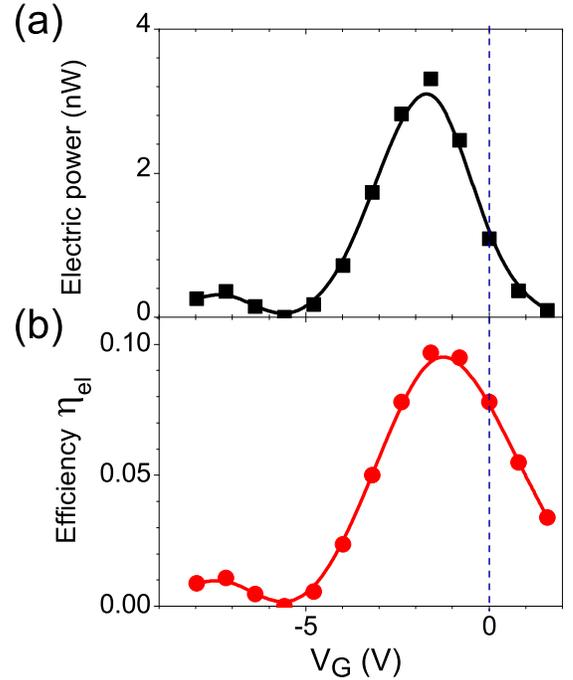}
\caption{
(color online)
(a) The electric power $\Delta P$, converted from the thermal energy using
the Seebeck effect, as functions of gate voltages. (b) the corresponding efficiency
of energy conversion $\eta_{el}$ vs. $V_{G}$. The temperatures of electrodes
are maintained at $T_{L}=200$~K and $T_{R}=300$~K}%
\label{Fig7}%
\end{figure}

The nanojunction can be considered as a field-effect transistor which is
powered by itself via the temperature difference maintained between electrodes
using the Seebeck effect. We observe that the gate voltages can shift the narrow
DOSs and transmission function near the chemical potential, as shown in
Fig.~\ref{Fig5}. Figure~\ref{Fig6} shows the induced potential $\Delta
V(T_{L},T_{R})$ [upper panel; given by Eq.~(\ref{DeltaVSeebeck1})] and current
$I(T_{L},T_{R})$ [lower panel; given by Eq.~(\ref{Ifinite2})] for
$T_{L}=200$~K and $T_{R}=300$~K as functions of gate voltages. We observe that
the extremal of $\Delta V(T_{L},T_{R})$ and $I(T_{L},T_{R})$ do not occurs
at the same value of $V_{G}$ because the gate voltage also modulates the conductance
of the junction [note that $\sigma=I(T_{L},T_{R})/\Delta V(T_{L},T_{R})$].
Figure~\ref{Fig6} demonstrates that the gate field can efficiently
control the magnitude, polarity, and power \textit{on-off} of the voltage and
current induced by the Seebeck effect. Such current-voltage characteristics could
be useful in the design of nanoscale electronic devices such as a transistor
or switch. Note that the induced voltage and induced current described in
Fig.~\ref{Fig6} comply with the sign convention described in Fig.~\ref{Fig1}(b).

Figure~\ref{Fig7} shows the electric power $\Delta P$ [upper panel; given by Eq.~(\ref{Power})]
delivered by the battery and the energy conversion efficiency
$\eta_{el}$ [lower panel; given by Eq.~(\ref{efficiency})] as functions of
the gate voltages. It illustrates that the gate field is capable of controlling
and optimizing the electric power and energy conversion efficiency $\eta_{el}%
$. The energy conversion efficiency $\eta_{el}\approx0.09$ if optimized by the
gate field. The optimized power is about $3.5$~nW at $V_{G}\approx-2$~V. We
also observe that the maximum values of electric power and energy conversion
efficiency occur at different voltages, similar to the case of induced voltage
and current. If a macroscopic system can be formed by two identical pieces of
surfaces, each of which is formed by a layer of Al atoms adsorbing onto metal
surface coated with a layer of Br atoms as spacer. The macroscopic system
could be equivalent to more than $10^{12}$ single paired metal-Br-Al-Al-Br-M
junctions which are connected in parallel on $1$~cm$^{2}$ surface. We
conjecture that the power could be substantial in such a macroscopic system.

\begin{figure}[ptb]
\includegraphics[width=8cm]{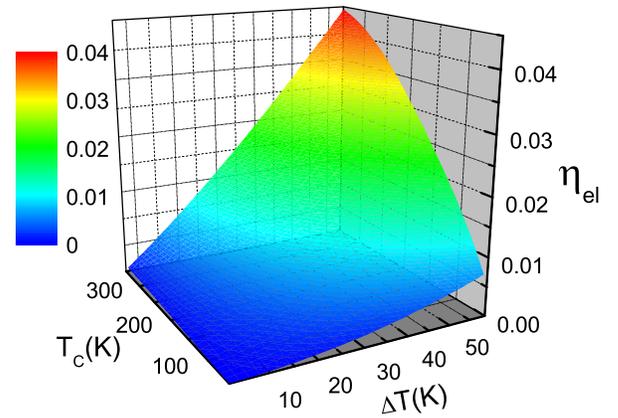}
\caption{
(color online)
Efficiency of energy conversion $\eta_{el}$ as a function of $T_{L}=T_{C}$ and
$\Delta T = T_{R}-T_{L}$.}
\label{Fig8}
\end{figure}

Finally, we investigate the efficiency of energy conversion $\eta_{el}$ as a
function of $T_{C}$ ($=T_{L}$) and $\Delta T=T_{R}-T_{L}$ for $V_{G}=0$ , as
shown in Fig.~\ref{Fig8}. The paired metal-Br-Al junction shows sufficiently
large efficiency around $0.05$ when $T_{C}=300$~K and $\Delta T=60$~K. Such a
high efficiency is attributed to the enhanced the Seebeck coefficients.

\section{Conclusions}

In summary, we developed a theory for thermoelectric power generator in the
truly atomic scale system. The theory is general to any atomic/molecular
junction where electron tunneling is the major transport mechanism. As an
example, we investigated the thermoelectric properties and the efficiency of
energy conversion of the paired metal-Br-Al junction from first-principles
approaches. Owing to the narrow states near the chemical potentials, the
nanojunction has large Seebeck coefficients; thus, it can be considered as an
efficient thermoelectric power generator. Provided that a finite temperature
difference is maintained between electrodes, the thermoelectric power
generator converts thermal energy into electric energy. The optimized electric
power generated by the Seebeck effect is about $3.5$~nW at $T_{L(R)}=200$
$(300)$~K at $V_{G}\approx-2$~V. To gain further insight into the mechanism of
energy conversion, we investigate the electron's thermal currents
analytically. The electron's thermal current which removes heat from the hot
temperature reservoir can be decomposed into two components,
$J_{el}^{R}=\left(J_{el}^{R}\right)  _{\Delta V}+\left(
J_{el}\right)  _{\Delta T}$. Only $\left(J_{el}^{R}\right)
_{\Delta V}$ is capable of converting energies. The electron's thermal current
removes heat from the hot reservoir via $\left(J_{el}^{R}\right)
_{\Delta V}$ and rejects waste heat into the cold reservoir via $\left(
J_{el}^{L}\right)  _{\Delta V}$. No energy conversion is possible when
the Seebeck coefficients is vanishing.

We also consider the nanojunction in a three-terminal geometry, where the
current, voltage, and electric power can be modulated by the gate voltages,
which shift the states of the junction. We observe that the gate field can
control the magnitude, power on-off, and polarity of the induced current and
voltages generated by the Seebeck effect. Such current-voltage characteristics
could be useful in the design of nanoscale electronic devices such as a
transistor or switch. Notably, the nanojunction as a transistor with a fixed
finite temperature difference between electrodes can power itself using the
Seebeck effect. The results of this study may be of interest to researchers
attempting to develop new forms of thermoelectric nano-devices.

The authors thank Ministry of Education, Aiming for Top University Plan (MOE
ATU), National Center for Theoretical Sciences (South), and National Science
Council (Taiwan) for support under Grants NSC 97-2112-M-009-011-MY3,
098-2811-M-009-021, and 97-2120-M-009-005.


\begin{thebibliography}{99}

\bibitem {Aviram}A.~Aviram and M.~A.~Ratner, Chem. Phys. Lett. \textbf{29}, 277 (1974).

\bibitem {Reed1}M.~A.~Reed, C.~Zhou, C.~J.~Muller, T.~P.~Burgin, and J.~M.~Tour, Science \textbf{278}, 252 (1997).

\bibitem {book}M.~Di~Ventra, \textit{Electrical transport in nanoscale systems}, (Cambridge University Press, Cambridge, 2008).

\bibitem {Kaun}C.~C.~Kaun and H.~Guo, Nano Lett. \textbf{3}, 1521 (2003).

\bibitem {DiVentra2}M.~Di~Ventra and N.~D.~Lang, Phys. Rev. B \textbf{65}, 045402 (2001).

\bibitem {Nitzan1}A.~Nitzan and M.~A.~Ratner, Science \textbf{300}, 1384 (2003).

\bibitem {Wang2}W.~Wang, T.~Lee, I.~Kretzschmar, and M.~A.~Read, Nano Lett. \textbf{4}, 643 (2004).

\bibitem {Ratner3}M.~Galperin, A.~Nitzan, and M.~A.~Ratner, Phys. Rev. B \textbf{78}, 125320 (2008).

\bibitem {Yluo}J.~Jiang, M.~Kula, W.~Lu, and Y.~Luo, Nano Lett. \textbf{5}, 1551 (2005).

\bibitem {Kushmerick}L.~H.~Yu, C.~D~Zangmeister, and J.~G.~Kushmerick, Phys. Rev. Lett. \textbf{98}, 206803 (2007).

\bibitem {Brandbyge}M.~Paulsson, T.~Frederiksen, and M.~Brandbyge, Nano Lett. \textbf{6}, 258 (2006).

\bibitem {Dicarlo}G.~C.~Slomon, A.~Gagliardi, A.~Pecchia, T.~Frauenheim, A.~Di~Carlo, J.~R.~Reimers, and N.~S.~Noel, J. Chem. Phys. \textbf{124},
094704 (2006).

\bibitem {alkane}J.~G.~Kushmerick and J.~Lazorcik, C.~H.~Patterson, and R.~Shashidhar, Nano Lett. \textbf{4}, 639 (2004).

\bibitem {cheniets}Y.~C.~Chen, M.~Zwolak, and M.~Di~Ventra, Nano Lett. \textbf{5}, 621 (2005).

\bibitem {chenh2}Y.~C.~Chen, Phys. Rev. B \textbf{78}, 233310 (2008).

\bibitem {thygesen}I.~S.~Kristensen, M.~Paulsson, K.~S.~Thygesen, and
K.~W.~Jacobsen, Phys. Rev. B \textbf{79}, 235411 (2009).

\bibitem {Ruitenbeek1}D.~Djukic and J.~M.~van~Ruitenbeek, Nano Lett. \textbf{6}, 789 (2006).

\bibitem {Kiguchi}M.~Kiguchi, O.~Tal, S.~Wohlthat, F.~Pauly, M.~Krieger, D.~Djukic, J.~C.~Cuevas, and J.~M.~van~Ruitenbeek, Phys. Rev. Lett.
\textbf{101}, 046801 (2008).

\bibitem {Natelson}P.~J.~Wheeler, J.~N.~Russom, K.~Evans, N.~S.~King, and D.~Natelson, Nano Lett. \textbf{10}, 1287 (2010).

\bibitem {chenshot}Y.~C.~Chen and M.~Di~Ventra, Phys. Rev. Lett. \textbf{95}, 166802 (2005).

\bibitem {liushot}Y.~S.~Liu and Y.~C.~Chen, Phys. Rev. B \textbf{83}, 035401 (2011).

\bibitem {chenheating}Y.~C.~Chen, M.~Zwolak, and M.~Di~Ventra, Nano Lett. \textbf{3}, 1691 (2003).

\bibitem {Huang2}Z.~Huang, B.~Xu, Y.~C.~Chen, M.~Di~Ventra, and N.~J.~Tao, Nano Lett. \textbf{6}, 1240 (2006).

\bibitem {DiVentragate}M.~Di~Ventra, S.~T.~Pantelides, and N.~D.~Lang, Appl. Phys. Lett. \textbf{76}, 3448 (2000).

\bibitem {Ma}C.~L.~Ma, D.~Nghiem, and Y.~C.~Chen, Appl. Phys. Lett. \textbf{93}, 222111 (2008).

\bibitem {Solomon1}P.~M.~Solomon and N.~D.~Lang, ACS Nano \textbf{2}, 435 (2008).

\bibitem {Solomon2}N.~D.~Lang and P.~M.~Solomon, ACS Nano \textbf{3}, 1437 (2009).

\bibitem {Reed}H.~Song, Y.~Kim, Y.~H.~Jang, H.~Jeong, M.~A.~Reed, and T.~Lee, Nature \textbf{462}, 1039 (2009).

\bibitem {Ahn}C.~H.~Ahn, A.~Bhattacharya, M.~Di~Ventra, J.~N.~Eckstein, C.~D.~Frisbie, M.~E.~Gershenson, A.~M.~Goldman, I.~H.~Inoue, J.~Mannhart,
A.~J.~Millis, A.~F.~Morpurgo, D.~Natelson, and J.~M.~Triscone, Rev. Mod. Phys. \textbf{78}, 1185 (2006).

\bibitem {Lindsay}S.~M.~Lindsay and M.~A.Ratner, Advanced Materials \textbf{19}, 23 (2007).

\bibitem {Tao1}N.~J.~Tao, Nat. Nanotechnol. \textbf{1}, 173 (2006).

\bibitem {Ruitenbeek}B. Ludoph and J.~M.~van~Ruitenbeek, Phys. Rev. B
\textbf{59}, 12290 (1999).

\bibitem {Reddy1}P.~Reddy, S.~Y.~Jang, R.~A.~Segalman, and A.~Majumdar,
Science \textbf{315}, 1568 (2007).

\bibitem {Reddy2}K.~Baheti, J.~A.~Malen, P.~Doak, P.~Reddy, S.~Y.~Jang, T.~D.~Tilley, A.~Majumdar, and R.~A.~Segalman, Nano Lett. \textbf{8}, 715 (2008).

\bibitem {Reddy3}J.~A.~Malen, P.~Doak, K.~Baheti, T.~D.~Tilley, R.~A.~Segalman, and A.~Majumdar, Nano Lett. \textbf{9}, 1164 (2009).

\bibitem {Malen}J.~A.~Malen, S.~K.~Yee, A.~Majumdar,and R.~A.~Swgalman, Chem. Phys. Lett. \textbf{491}, 109 (2010).

\bibitem {Paulsson}M.~Paulsson and S.~Datta, Phys. Rev. B \textbf{67}, 241403(R) (2003).

\bibitem {Zheng}X.~Zheng, W.~Zheng, Y.~Wei, Z.~Zeng, and J.~Wang, J. Chem. Phys. \textbf{121}, 8537 (2004).

\bibitem {Wang}B.~Wang, Y.~Xing, L.~Wan, Y.~Wei, and J.~Wang, Phys. Rev. B \textbf{71}, 233406 (2005).

\bibitem {Pauly}F.~Pauly, J.~K.~Viljas, and J.~C.~Cuevas, Phys. Rev. B \textbf{78}, 035315 (2008).

\bibitem {Dubi}Y.~Dubi and M.~Di~Ventra, Nano Lett. \textbf{9}, 97 (2009).

\bibitem {Markussen}T.~Markussen, A.~P.~Jauho~, and M.~Brandbyge, Phys. Rev. Lett. \textbf{103}, 055502 (2009).

\bibitem {Ke}S.~H.~Ke, W.~Yang, S.~Curtarolo, and H.~U.~Baranger, Nano Lett. \textbf{9}, 1011 (2009).

\bibitem {Finch}C.~M.~Finch, V.~M.~Garc\'{\i}a-Su\'{a}rez, and C.~J.~Lambert, Phys. Rev. B, \textbf{79}, 033405 (2009).

\bibitem {Troels}T.~Markussen, A.~P.~Jauho, and M.~Brandbyge, Phys. Rev. B, \textbf{79}, 035415 (2009).

\bibitem {Bergfield}J.~P.~Bergfield and C.~A.~Stafford, Nano Lett. \textbf{9}, 3072 (2009).

\bibitem {Liu2}Y.~S.~Liu, Y.~R.~Chen, and Y.~C.~Chen, ACS Nano \textbf{3}, 3497 (2009).

\bibitem {Galperin}M.~Galperin, A.~Nitzan, and M.~A.~Ratner, Molecular Phys. \textbf{106}, 397 (2008).

\bibitem {Imry}O.~Entin-Wohlman, Y.~Imry, and A.~Aharony, Phys. Rev. B \textbf{82}, 115314 (2010).

\bibitem {Hsu}B.~C.~Hsu, Y.~S.~Liu, S.~H.~Lin, and Y.~C.~Chen, Phys. Rev. B (in press).

\bibitem {Esposito}M.~Esposito, K.~Lindenberg and C.~Van den Broeck, Phys. Rev. Lett. \textbf{102}, 130602 (2009).

\bibitem {Liu}Y.~S.~Liu and Y.~C.~Chen, Phys. Rev. B \textbf{79}, 193101 (2009).

\bibitem {DiventraS}Y.~Dubi and M.~Di~Ventra, Phys. Rev. B \textbf{79}, 081302(R) (2009).

\bibitem {Galperin2}M.~Galperin, K.~Saito, A.~V.~Balatsky, and A.~Nitzan, Phys. Rev. B \textbf{80}, 115427 (2009).

\bibitem {Liu3}Y.~S.~Liu, B.~C.~Hsu, and Y.~C.~Chen,~\textit{cond-mat/arXiv:0908.0992}.

\bibitem {Heinrich}C.~F.~Hirjibehedin, C.~P.~Lutz, and A.~J.~Heinrich, Science \textbf{312}, 1021 (2006).

\bibitem {Lang2}N.~D.~Lang, Phys. Rev. B \textbf{55}, 9364 (1997).

\bibitem {Lyo}I.~M.~Lyo, Ph.~Avouris, Science \textbf{245}, 1369 (1989).

\bibitem {Lang3}M.~L.~Yu, N.~D.~Lang, B.~W.~Hussey, T.~H.~P.~Chang, and W.~A.~Mackie, Phys. Rev. Lett. \textbf{77}, 1636 (1996).

\bibitem {Justin}J.~P.~Bergfield, M.~A.~Solis, and C.~A.~Stafford, ACS Nano \textbf{4}, 5314 (2010).

\bibitem {JianWang}J.~Wang and H.~Guo, Phys. Rev. B, \textbf{79}, 045119 (2009).

\bibitem {Lang4}N.~D.~Lang, Phys. Rev. B, \textbf{45}, 13599 (1992).

\bibitem {Lang}N.~D.~Lang, Phys. Rev. B \textbf{52}, 5335 (1995).

\bibitem {Chen}Y.~C.~Chen and M.~Di~Ventra, Phys. Rev. B \textbf{67},153304 (2003).
\end{thebibliography}
\end{document}